\documentclass[preprint,aps,prd,nofootinbib,superscriptaddress]{revtex4}

\usepackage{graphicx,fancyhdr,lineno,multirow,rotating,calc,amssymb}

\usepackage{epsfig}
\usepackage{epsf}

\def\Real#1{\Re\left\{#1\right\}}

\def\beq{\begin{equation}}
\def\eeq{\end{equation}}

\def\beqa{\begin{eqnarray}}
\def\eeqa{\end{eqnarray}}

\def\pzq{{\ensuremath{\pi^0\pi^0\pi^0}}}

\def\etapzq{\ensuremath{\eta\to\pzq}}

\def\matrel{\ensuremath{{\cal M}}}
\def\matpppz{\ensuremath{{\matrel}_{+-0}}}
\def\matpzq{\ensuremath{{\matrel}_{000}}}
\def\mats{\ensuremath{{\matrel}_S}}

\def\matvp{\ensuremath{{\matrel}_{\rho^+}}}
\def\matvm{\ensuremath{{\matrel}_{\rho^-}}}

\def\matvmp{\ensuremath{{\matrel}_{\rho^\mp}}}
\def\matrho{\ensuremath{{\matrel}_\rho}}

\def\babar{\mbox{\slshape B\kern-0.1em{\small A}\kern-0.1em
    B\kern-0.1em{\small A\kern-0.2em R}}}

\def\Getarhopi{\ensuremath{g_{\eta\rho\pi}}}
\def\Grhopipi{\ensuremath{g_{\rho\pi\pi}}}

\def\BR{{\ensuremath{\cal B}}}

\def\channel{\ensuremath{\tau^-\to\pi^-\eta\nu_\tau}}

\def\ket|#1>{\left|#1 \right>}

\def\bra<#1|{\left< #1 \right|}

\def\bracket<#1|#2>{\setbox0=\vbox{\hbox{$#1$$#2$}}\left<#1\kern1pt \vrule  height\ht0\kern2pt #2\right>}

\def\dirmat<#1|#2|#3>{\setbox0=\vbox{\hbox{$#1$$#2$$#3$}}\left<#1\kern1pt \vrule height\ht0\kern1pt#2\kern1pt \vrule height\ht0\kern1pt #3\right>}

\begin{document}

\title{\large Estimates of the branching fraction \boldmath 
     $\tau^- \to \eta \pi^-\nu_\tau$,  
     the $a_0^-(980)$, and non-standard weak interactions}
\bigskip

\author{S.~Nussinov}
\affiliation{Tel Aviv University, Tel Aviv, 69978, Israel }
\affiliation{University of Maryland, College Park, MD 20742, USA}
\author{A.~Soffer}
\affiliation{Tel Aviv University, Tel Aviv, 69978, Israel }

\date{\today}

\bigskip
\bigskip

\begin{abstract}
We consider the ``second-class current'' decay \channel\ from several
points of view. We first focus on the decay rate as expected within
standard weak interaction and QCD due to isospin violation.  The decay
contributions divide into $P$- and $S$-wave parts. The former can be
reliably estimated using the $\rho\eta\pi$ coupling inferred from the
rates and Dalitz-plot distributions of $\eta\to 3\pi$ decays.  The
somewhat larger $S$-wave part, which was previously computed using
chiral perturbation theory, is estimated from a simple $\bar qq$
model.  Both estimates of the $S$-wave part depend on whether the
$a_0(980)$ scalar particle is a $\bar qq$ or some other (4-quark)
state. Finally, we discuss genuinely new, non-$V-A$
scalar weak interactions. The \channel\ decay provides information on
this question, which nicely complements that from precision $\beta$
decay experiments.
In summary, we discuss the possible implications of putative values of
the branching fraction $\BR(\channel)$. In the case of larger values,
in particular of the $S$-wave part, not only will detection of the
decay be more likely and more reliable, its implications will be
more far-reaching and interesting.

\end{abstract}

\maketitle

\bigskip
\bigskip

\section{Introduction}
The weak decay \channel, an example of ``second class current'' decays
introduced by Weinberg~\cite{ref:weinberg}, may
soon be observed or tightly bounded by the B~factories.  This isospin-
and G parity-violating decay is suppressed by the small value of
$(m_d - m_u)/\Lambda_{QCD}$ or $\alpha_{EM}$.
Various estimates~\cite{ref:oldpred} using chiral perturbation theory
or other methods have predicted this decay's branching fraction to be
\beq
  \BR \equiv \BR(\channel) = (1.3 \pm 0.2) \times 10^{-5},
\label{eq:oldpred}
\eeq
far below the present CLEO upper bound of 
$1.4\times 10^{-4}$~\cite{Bartelt:1996iv}.
In view of the possibility of new measurements, we point out interesting consequences
of various \BR\ values.

The plan of this paper is as follows:
In Sec.~\ref{sec:kin} we present the kinematics and some other general
aspects of the $\channel$.
The contribution of the vector ($L=1$) $\pi^-\eta$ final state to \BR\
is discussed in Sec.~\ref{sec:vector}, assuming that the $L=1$ and
$I=1$, $\pi^-\eta$ final state is dominated by the $\rho^-$ meson.
Sec.~\ref{sec:IV} addresses the contribution of the $J^P = 0^+$
$\pi^-\eta$ state to \BR.  The analog of the $\rho^-$ here is the
$I=1$, $a_0^-(980)$ state, whose coupling to the vector current relates
to a longstanding question on whether the $a^-_0(980)$ is a $\bar ud$
state or a $\bar ud\bar ss$/$\bar KK$-threshold state.
The $\bar ud$ assumption was implicitly made in the
chiral-Langrangian calculations predicting Eq.~(\ref{eq:oldpred}),
where $a^-_0(980)$ dominance was used to analytically continue
the calculation of low-energy decays to the $\tau^-$ decay of interest.  We
briefly discuss another naive quark-model-based estimate. 
Sec.~\ref{sec:V} addresses the possible relation between \BR\ and
precise measurements of $\beta$-decay spectra from trapped radioactive
ions. Such measurements can be used to search for
scalar interactions, in addition to the standard electroweak
$(V-A)\cdot (V-A)$ interaction.
In the concluding Sec.~\ref{sec:VI} we present putative \BR\ values
and/or bounds on \BR\ with implications for the discussions in the
former sections.

\section{Kinematics of the \boldmath $\channel$ Decay}
\label{sec:kin}

Only the vector weak current
$ V_{\mu}(x)= \bar u(x) \gamma_{\mu}d(x) $
contributes to the hadronic part
$ \dirmat<0|J^W_{\mu}| \eta \pi^-> = H_{\mu} $
of the current-current interaction, since  the $1^+$ and $0^-$ parts of the
axial current cannot create natural-parity states of two
pseudoscalars.  The matrix element $H_{\mu}$ can be decomposed into a
$J^P =0^+$ part and a $1^-$ part in the rest frame of the $\eta \pi^-$
system as follows:
\beq
\dirmat<0|V_{\mu}|\pi\eta> = 
    f_1(s) q_\mu
  + f_0(s) Q_\mu,
\label{eq1}
\eeq
where $f_L$ is the coefficient of the state with angular-momentum $L$,
\beqa
Q &\equiv& q_\pi + q_\eta, \nonumber\\
q &\equiv& a(s)q_\pi - q_\eta, \nonumber\\
s &\equiv& Q^2,
\label{eq:Qqs}
\eeqa
$q_x$ is the four-momentum of particle $x$,
and
\beq
 a(s) \equiv  {m_{\eta}^2 + q_1 \cdot q_2 \over m_{\pi}^2 + q_1 \cdot q_2} 
\label{eq2}
\eeq
is chosen so that $Q \cdot q=0$.
In the rest frame of the $\eta\pi^-$ system, $q$ is a space-like vector:
\beq
 q =(0,|q| \cos\theta , |q| \sin\theta ,0),
\label{eq3}
\eeq
where $\theta$ is the angle in this frame between $\vec q$ and the
recoiling neutrino momentum.
The $L=0$ and $L=1$ amplitudes interfere in the
angular dependence $d\Gamma/d(cos(\theta)$, but not in the total
decay rate obtained by integrating over $d(cos(\theta)$, namely,
\beq
 {d\Gamma \over ds} =  K_1|f_1(s)|^2 + K_0|f_0(s)|^2,
\label{eq4}
\eeq
with the $K_L$ being kinematic factors. 
Thus, either the $S$- or $P$-wave contribution yields a lower bound 
on the total rate. We proceed with an estimation
of the magnitudes of these contributions.
%

\section{Estimating the \boldmath  $L=1$ Contribution of the $\pi^-\eta$ State}
\label{sec:vector}

The decay $\tau^-\to \pi^-\pi^0\nu_\tau$ comprises $25.5\%$ of all $\tau^-$
decays, and is completely dominated by $\rho^-$ exchange. 
Similarly, our estimate of the $L=1$ contribution to the
decay \channel\ assumes $\rho^-$ dominance, taking place via 
$\tau^-\to \rho^-\nu_\tau$ followed by 
$\rho^-\to\eta\pi^-$. We thus expect the $L=1$ component of $\BR$
to be
\beq
\BR_{L=1} = \left({\Getarhopi \over \Grhopipi}\right)^2 
           \left({p_{\rho\to\eta\pi} \over p_{\rho\to\pi\pi}}\right)^3  
           \BR(\tau^-\to \rho^-\nu_\tau),
\label{eq:BRL=1.1}
\eeq
where \Getarhopi\ and \Grhopipi\ are the $\rho\to\eta\pi$ and
$\rho\to\pi\pi$ coupling constants, respectively, and the cubed ratio
between the daughter momenta in the two decays is
$\left(p_{\rho\to\eta\pi} / p_{\rho\to\pi\pi}\right)^3 = 0.07$.

Since the decay $\rho^-\to\eta\pi^-$ has not been observed, 
we obtain the coupling constant \Getarhopi\ from
the Dalitz-plot distribution of the decay 
$\eta\to\pi^+\pi^-\pi^0$ and the branching fraction
$\BR(\eta\to\pi^0\pi^0\pi^0)$. The three-pion Dalitz plot is 
customarily described with the variables 
\beqa
X &\equiv& {\sqrt{3} \over Q} (T_+ - T_-), \nonumber\\
Y &\equiv& {3 \over Q} T_0 - 1,
\label{eq:xy}
\eeqa
where $T_c$ is the kinetic energy of the pion with charge $c$, and 
\beq
Q \equiv m_\eta - 2m_{\pi^+} - m_{\pi^0} \approx  m_\eta - 3 m_\pi.
\label{eq:Q}
\eeq
Henceforth, we ignore the difference between the charged and
neutral pion masses.
The matrix element for $\eta\to\pi^+\pi^-\pi^0$
is taken to be the sum of a scalar and a vector exchange contribution,
the latter dominated by the $\rho(770)$:
\beq
\matpppz = \mats + \matvp + \matvm.
\label{eq:matpppz}
\eeq
A $\rho^0$ contribution is forbidden due to charge conjugation conservation. 
Properly accounting for the number of diagrams and identical particles, 
the \etapzq\ matrix element is
\beq
\matpzq = {3 \over \sqrt{3!}} \mats.
\label{eq:matpzq}
\eeq
The branching fraction of this decay gives the absolute value of the
scalar matrix element,
\beq
|\mats|^2 = 8 (2\pi)^3 m_\eta  \Gamma_\eta  \BR(\eta\to\pi^0\pi^0\pi^0) 
                 {6 \sqrt{3} \over Q^2 S_1} {3! \over 9} = 0.065,
\label{eq:mats}
\eeq
where we used the measured values of the $\eta$ mass, width, and
$\pi^0\pi^0\pi^0$ branching fraction~\cite{ref:pdg06}, the phase-space
differential is $dE_1 dE_2 = (Q^2 / 6 \sqrt{3}) dX\, dY$, and $S_1 =
2.75$ is the area of the Dalitz plot. The scalar particle exchanged is
assumed to be very broad, so that the distribution of events over the
relatively small Dalitz plot is essentially uniform.

We take the vector matrix element to be
\beqa
\matvmp &=& -\Getarhopi \Grhopipi {(P_\eta + P_\pm) \cdot (P_\mp - P_0)
          \over
          (P_\mp + P_0)^2 - m_\rho^2 - i\Gamma_\rho m_\rho
         } \nonumber \\
   &=& 
    -\Getarhopi \Grhopipi 
      {2 m_\eta  \left(E_0 - E_\mp \right) \over 
                2 m_\eta E_\pm + M_0^2 - {2 \over 3} m_\eta^2}, 
\label{eq:matvm1}
\eeqa
where $E_+$, $E_-$, and $E_0$ are the $\eta$-rest-frame energies 
of the $\pi^+$, $\pi^-$, and $\pi^0$, respectively, and
\beq
M_0^2 \equiv m_\rho^2 - {1 \over 3} m_\eta^2 - m_\pi^2 + i\Gamma_\rho m_\rho.
\label{eq:defM2}
\eeq
%
Replacing the energies with the Dalitz-plot quantities of 
Eqs.~(\ref{eq:xy}) and~(\ref{eq:Q}),
the sum of the $\rho^+$ and $\rho^-$ contributions is
\beqa
\matvm + \matvp &=&  -2 \Getarhopi \Grhopipi 
 {
    r Y -\frac13 r^2 (Y^2 + X^2)
     \over 1 - \frac23 r Y + \frac13 r^2(\frac13 Y^2 - X^2)
 } \nonumber\\
&\approx&
  -\Getarhopi \Grhopipi 2
    \left[ r Y + {r^2\over 3} \left(Y^2 - X^2 \right)
      + {r^3 \over 9}\left(X^2Y -Y^3 \right)
    \right],  
\label{eq:matv2}
\eeqa
where 
\beq
r \equiv {m_\eta Q \over M_0^2} = 0.14 + 0.03i.
\eeq
and the last line of Eq.~(\ref{eq:matv2}) is obtained from a Taylor 
expansion to order $r^3$.

Squaring the sum of the scalar and vector terms, again keeping 
terms to order $r^3$, we obtain
\beq
{|\matpppz|^2 \over |\mats|^2} \approx
1 + \alpha Y + \beta Y^2 + \gamma X^2 + \delta Y^3 - \delta YX^2,
\label{eq:dpdist}
\eeq
where
\beqa
\alpha &=& -4 \Getarhopi \Grhopipi \Real{\mats^* r} {1\over |\mats|^2} ,
     \nonumber\\
\beta &=& \left[-\frac43 \Getarhopi \Grhopipi \Real{\mats^* r^2} 
          +  4(\Getarhopi \Grhopipi)^2 |r|^2 \right] {1\over |\mats|^2}  ,
    \nonumber\\
\gamma &=& \frac43 \Getarhopi \Grhopipi \Real{\mats^* r^2} {1\over |\mats|^2} ,
    \nonumber\\
\delta &=& \left[ \frac49 \Getarhopi \Grhopipi \Real{\mats^* r^3} 
      + \frac83 (\Getarhopi \Grhopipi)^2 \Real{r (r^2)^*}\right] 
            {1\over |\mats|^2} .
\label{eq:coeffs}
\eeqa

The product of coupling constants $\Getarhopi \Grhopipi$ is obtained
by comparing the coefficients of Eq.~(\ref{eq:dpdist}) with the
Dalitz-plot distribution of the decay $\eta\to\pi^+\pi^-\pi^0$. 
A high-statistics study of this distribution has been
recently performed by the KLOE collaboration~\cite{ref:kloe},
yielding the parameterization
\beq
|\matpppz|^2 \propto 1 -1.09 Y + 0.124 Y^2 + 0.057 X^2 + 0.14 Y^3.
\label{eq:dp-exp}
\eeq
We ignore the measured coefficient errors, as they are much smaller
than the theoretical errors associated with our model. From the
coefficient of the $Y$ term in Eq.~(\ref{eq:dp-exp}) and the first of
Eqs.~(\ref{eq:coeffs}), one obtains the product of coupling constants
\beq
\Getarhopi \Grhopipi =  {1.09 \over 4} {\mats \over \Real{r} } = 0.51,
\label{eq:ggFromY}
\eeq
where $\mats$ was taken to be real.
The accuracy of the model may be judged from the values it obtains 
for the other coefficients: 
\beq
|\matpppz|^2 \propto 1 -1.09 Y + 0.27 Y^2 + 0.05 X^2 + 0.03 Y^3 - 0.03 YX^2.
\label{eq:dp-model}
\eeq
Allowing $\mats$ to have a complex phase does not improve the agreement
between Eqs.~(\ref{eq:dp-exp}) and~(\ref{eq:dp-model}) significantly. 
A related cross-check is provided by the ratio of branching fractions
$\BR(\eta\to\pi^+\pi^-\pi^0)/\BR(\eta\to\pi^0\pi^0\pi^0) = 0.70$.
The value predicted by Eqs.~(\ref{eq:dpdist}) and~(\ref{eq:mats})
is 0.71 when using the experimental coefficients of Eq.~(\ref{eq:dp-exp}),
and 0.76 using those of Eq.~(\ref{eq:dp-model}).

Taking the matrix element for the decay $\rho\to\pi\pi$ to be 
\beq 
\matrho = \Grhopipi \varepsilon_\mu^{(\xi)} (P_+ - P_-)^\mu,
\eeq
the coupling constant \Grhopipi\ is determined to be
\beq
\Grhopipi = \sqrt{
                    6 \pi m_\rho^2 \Gamma_\rho \over p_{\rho\to\pi\pi}^3
                 }
          = 6.0.
\label{eq:grhopipi}
\eeq
Eqs.~(\ref{eq:grhopipi}) and~(\ref{eq:ggFromY}) then give
\beq
\Getarhopi \approx 0.085.
\label{eq:Getarhopi}
\eeq
A similar calculation by Ametller and
Bramon~\cite{ref:ametller} yielded the ratio 
$\Getarhopi / \Grhopipi = 0.011 \pm 0.002$, consistent with
our results.

From Eqs.~(\ref{eq:BRL=1.1}), (\ref{eq:grhopipi}), and~(\ref{eq:Getarhopi})
we calculate the $L=1$ component of the \channel\ branching fraction,
\beq
\BR_{L=1} \approx 3.6 \times 10^{-6}. 
\label{eq:BRL=1.2}
\eeq
We also obtain 
\beq
\BR(\rho\to\eta\pi) = {\Getarhopi^2 p_{\rho\to\eta\pi}^3 
  \over 6 \pi m_\rho^2 \Gamma_\rho} \approx 1.4 \times 10^{-5},
\eeq
far below the current experimental limit of $6\times 10^{-3}$~\cite{ref:pdg06}.

\section{The \boldmath $L=0$ Contribution}
\label{sec:IV}

The Contribution of the ($L=0$) $\pi^-\eta$ state to \BR\ is not as
readily accessible to a phenomenological estimate as that of the $L=1$
state. The observed $\rho^-$ dominance in the $\pi^- \pi^0$
final state of the $\tau^-$ decay is expected, since the $\rho^-$ has
the quantum numbers of the hadronic vector current $\overline u
\gamma_\mu d$. It is therefore natural to assume that it also
dominates the ($L=1$) $\pi^-\eta$ final state, although this decay is
suppressed by isospin violation.
This is not so for the superficially analog case of $a^-_0(980)$ and
the scalar contribution to \BR. In Ref.~\cite{ref:oldpred}, 
the $a^-_0(980)$ dominance of the ($L=0$)
$\pi^-\eta$ channel in weak decays was used to extrapolate the
low-energy amplitude for $\eta \to \pi^- e^+ \nu_e$ (computed via
chiral perturbation theory) to the decay
$\channel$ and obtain the estimate of
Eq.~(\ref{eq:oldpred}). 
The resulting scalar contribution to \BR\ is then $\sim 3$ times
larger than the vector contribution.  This extrapolation is
questionable not only because of the large change in $Q^2$ from
$\sim 0.15$~GeV$^2$ to $\sim 1$~GeV$^2$.  The key point is that
$a^-_0(980)$ (just like its $I=0$ counterpart $f_0(980)$) may well be
a four-quark $\bar ud\bar ss$ state, a view suggested early
on~\cite{ref:jaffe} and adopted recently by the Particle Data
Group~\cite{ref:pdg06}. In this case, the $a_0(980)$ coupling to the
$\bar ud$ scalar current is ``Zweig-Rule'' suppressed, and the
four-quark state will not dominate the decay in question.

Several considerations suggest that the $a_0(980)$ and $f_0(980)$ states
have significant four-quark contributions: 
\begin{enumerate}
\item \label{arg1} The widths $\Gamma(f_0(980)\to \pi\pi) \sim \Gamma
(a_0(980)\to\pi \eta) \sim 50$~MeV are anomalously small for an S-wave
$\bar qq$ state. Since the lighter, 770-MeV $\rho$ has a
P-wave decay width of 150~MeV, the $a_0(980)$ $f_0(980)$ and widths
should have been vastly larger. This is the case for the
so-called $\sigma(600)$ scalar, often used in nuclear potentials, which has 
a width of about 600~MeV.
\item \label{arg2} The fact that $a_0(980)$ and $f_0(980)$ decay also
into $K\bar K$ despite the highly reduced phase space (the decay is
kinematically forbidden over most of the widths) is an argument against
their being $\bar qq$ states. Indeed, four-quark states would much more
readily fall apart to $q\bar s \bar qs = \bar KK$ than would $\bar q
q$ scalars. 
In principle, the $a_0$ and $f_0$ could be 
"molecular", lightly bound $\bar KK$ threshold states, 
in analogy with the $X(3872)$, which may be a $D^*\bar D$ threshold 
state~\cite{ref:3872}.
For states of similar size, the kinetic energy in the $D^*\bar D$
system is four times smaller than that of the $\bar KK$ system. On the
other hand, roughly the same meson-meson potentials are generated by
couplings of the light quarks.  Therefore, binding $\bar KK$ to form
$a_0(980)$ and $f_0(980)$ seems unlikely.  
The features~\ref{arg1} and~\ref{arg2} above, which are particularly
puzzling in a $\bar qq$ picture, can conceivably be resolved if one
notes the special role of t'Hooft's anomaly-induced $\bar u u\bar d d
\bar s s$ six-quark coupling~\cite{Hooft:2008we}.

\item\label{arg4} Further indirect support for 
the four-quark picture comes from the
suggestion~\cite{nussinov:2004ud} that in collision or decay processes
with few initial quarks, $\bar qq$ meson production should exceed
considerably that of more complex baryonic and exotic four-quark
states. Comparison of $a_0(980)$ and $f_0(980)$ with
bonafide $\bar qq$ states such as $\rho (770)$ mesons in $e^+e^-$
or $p\pi$  collisions and in $B$ decays suggests that the former are
significantly suppressed, again supporting the four-quark hypothesis.
If the initial state has many quarks and, in particular, many $\bar s
s$ pairs, as is the case at the Relativistic Heavy Ion Collider, then
the suppression of $\bar qq\bar ss$ production is expected to be
weaker. This may be easier to test for $f_0(980)$ than for $a_0(980)$,
whose identification requires good photon reconstruction.
As further example, we note that 11\% of the decay $D_s^+\to K^+K^-\pi$
is due to $f_0\to K^+K^-$~\cite{ref:pdg06}.
\end{enumerate}

If $a_0(980)$ is indeed a four-quark state, then \BR\ 
will be smaller than the value predicted utilizing $a_0(980)$
dominance and assuming it is a $\bar qq$ state, Eq.~(\ref{eq:oldpred}).
If a search for $\channel$ that is sensitive to a branching fraction of order
$10^{-5}$ fails to detect a $\sim 50$~MeV-wide peak around 980~MeV in
the $\eta \pi^-$ invariant mass spectrum, this would constitute a
fourth argument in support of the four-quark view.
Conversely, observation of a clear peak would strongly suggest that
$a_0(980)$ is in fact a regular $\bar ud$ state, as early arguments by
Bramon and Masso have suggested~\cite{ref:bramon-masso}.

Next, we present some general arguments regarding the expected scalar
($L=0$) contribution $\BR_S$ to the branching fraction \BR,
assuming that it is dominated by the exchange of the 
$a_0(980)$, which is taken to be a $\bar u d$ state.  Key to
its small magnitude is the operator equation expressing the fact that
the weak vector current is conserved up to small electromagnetic and
$m_d - m_u$ mass difference effects:
\beq
\nabla^\mu V_{\mu }(x)=(m_d - m_u) \bar u(x)d(x) 
        + eA_{em}^{\mu }(x)V_{\mu }(x).
\label{eq:*}
\eeq
The contribution of the electromagnetic interaction term to
$\tau\to\pi\eta\nu_\tau$ is related to $\tau\to\pi\eta\nu_\tau\gamma$,
but given the difficulty in observing $\tau\to\pi\eta\nu_\tau$,
there is little hope that the decay involving an additional
photon in the final state can be studied in the near future.
The corresponding one-loop electromagnetic corrections are supressed
by $\alpha/\pi \sim 1/500$. The first term of Eq.~(\ref{eq:*}) is
$\sim(m_d-m_u)/ m_h \sim 1/200$ for $(m_d-m_u) \sim 4$~MeV and
a typical hadronic mass of $m_h \sim 0.8$~GeV, hence we focus on this
term in what follows.
The matrix element $\dirmat<0|\nabla^\mu V_{\mu }|h>$ (with
$h=\eta\pi$ or $h=a_0^-(980)$, if $a_0^-(980)$ dominance holds) of the
operator equation~(\ref{eq:*}) then yields
\beq
 Q^{\mu } \dirmat<0|V_{\mu}|h> = Q^2 f_0(s)= 
           (m_d - m_u) \dirmat<0|S^-|h>,
\label{eq:**}
\eeq
where $S^-$ is the scalar current $\bar u(x)d(x)$ , and $Q^2=s=m_h^2$
is the squared mass of the hadronic system.  The left-hand side 
of Eq.~(\ref{eq:**}) yields
the middle expression by using Eqs.~(\ref{eq1}) and~(\ref{eq:Qqs}).
Thus, computing $\BR_{L=0}$, the $L=0$ contribution to \BR, reduces to
estimating the low-energy hadronic parameter
$\dirmat<0|S^-|h>$.  A first-principles, unquenched lattice QCD
calculation is lacking at present, but recent progress in dealing with
light quarks/pseudoscalars may soon make it
feasible~\cite{ref:bernard}. The calculation is circumvented in the
chiral perturbation theory approach, which uses effective Lagrangians
(including isospin violation) and couplings fitted together to known
low-energy processes and extrapolated to the $\tau$ decay of
interest. The fact that as many as three calculations of this type
yielded the same result (Eq.~(\ref{eq:oldpred})) indicates that this
is a well-defined framework, but does not test its reliability.

Here we present a simpler quark model-motivated estimate. 
Unlike the $A\sim V$ and $S\sim P$ chiral 
symmetry-motivated relation, we relate the axial and scalar matrix
elements, since both pertain to $P$-wave ($a_1(1260)$ and $a_0(980)$) rather
than $S$-wave ($\rho$ and $\pi$) $\bar q q$  states.
We assume that
$a_0^-(980)$ dominates the \channel\ decay and that the decay
$\tau\to \pi^-\pi^+\pi^-\nu_\tau$ is dominated by the $a_1^-(1260)$.
Defining the matrix elements 
\beqa
v &\equiv& \dirmat<0|S^- |a_0^-(980)>,\nonumber\\
a &\equiv& \dirmat<0|A_i| a_1^-(1260)_i>,
\eeqa
where $i$ is a helicity state index, we expect
\beq
{\BR_{L=0} \over \BR(\tau^-\to a_1^-(1260)\nu_{\tau})} 
           \sim 1.3 {v^2 \over a^2} {\left(m_d - m_u \over m_{a_1(1260)} 
                 \right)^2},
\label{eq:a0a1-ratio}
\eeq
where the 1.3 enhancement is due to the larger phase space for the decay into
the lighter $a_0^-(980)$.
The couplings of the local scalar and axial currents to the two $^3P_0$ and
$^3P_1$ $\bar u d$ states of similar mass are expected to
be roughly equal, namely, $a \sim v$. Indeed, these couplings are fixed by
quark-model wave functions which, apart from relatively small $L \cdot
S$ effects, are the P-wave ground states of the same Hamiltonian. 
From Eq.(\ref{eq:a0a1-ratio}) we find
\beq
\BR_{L=0} \sim 1 \times 10^{-5},
\label{eq:L=0-est}
\eeq
similar to the contribution of the $\rho^-$ and, within our
crude approximations, consistent with the chiral-perturbation-theory
estimates. 
We note that Eq.~(\ref{eq:L=0-est}) may require an additional
suppression factor of up to $\sim3$, due to the three helicity states
available to the $a_1^-(1260))$.

\section{Test for New Weak Interactions}
\label{sec:V}

The general Lorentz-invariant "current $\times$ current" weak
interactions could include, in addition to $(V-A)\cdot(V-A)$, 
products of scalar ($S$), pseudoscalar ($P$) and tensor ($T$)
"currents".  Exchanging new, heavy elementary particles cannot
generate the non-minimal $T$ part, hence we focus on the $S$ and $P$
parts. 
Experimentally, the amplitudes of the $V\cdot V$, $V\cdot A$, and
$A\cdot A$ current products can be compared with those of $S\cdot
S$, $P\cdot P$ or $S\cdot P$ terms in nuclear beta decays involving
both $u \to d$ and $e\to\nu_e$ weak transitions~\cite{Gorelov:2004hv}.
It is convenient to parameterize the corrections to the Standard-Model
currents using the same weak coupling $g _W^2$,
attributing the smallness of the $S \cdot S$, $S\cdot P$ and $P\cdot
P$ terms to heavy (pseudo-)~scalar mesons with masses $m_P , m_S \gg
m_W$.  A positive result implying $m_S, m_P$ masses smaller than
$O$(TeV) would motivate searching for such particles at the upcoming
LHC.

A stringent limit on the pseudoscalar mass $M_P$ comes from its contribution
of $g_W^2/{M_P^2}$ to the amplitude $A(\pi^- \to e^- \nu_e)$. The
branching fraction for this decay, $(1.230 \pm 0.004)\times 10^{-4}$,
is in agreement with the expectation of the standard electroweak
model, where its small value is due to the $m_e / m_{\mu } \sim 1/200$
helicity suppression of the $V-A$ amplitude.  We therefore use the
error of this result to obtain an approximate limit on the
pseudoscalar contribution,
\beq
\left({M_W \over M_P}\right)^2 < 0.004 \times 10^{-4}  {1 \over 200} 
   \sim 3 \times 10^{-6}.  
\eeq
In order for measurements using unsuppressed nuclear beta decays to
compete with this limit, a precision of about $3 \times 10^{-6}$ is
needed.   
Similarly, the decays $K^- \to e^-\bar \nu_e$ and $B^- \to
e^-\bar\nu_e$ yield stringent bounds on pseudoscalar couplings
involving second- and third-generation 
quarks~\cite{ref:hou}.
We note that direct production of a pseudoscalar with mass $M_P > 10^3
M_W$ is far beyond the reach of the LHC.

The case of the scalar part is different. Current limits from
high-precision nuclear beta-decay experiments will continue to be
unchallenged by accelerator-based experiments, until an eventual
$B$-factory limit on or observation of the decay \channel,
whose small Standard-Model branching fraction makes it sensitive
to new scalar interactions.
In a nuclear beta decay, the distribution of the angle beween the
neutrino and the lepton is
\beq
W(\theta)= 1 + b {m_e \over E_e} + a \beta_e \cos(\theta),
\eeq
where $m_e$, $E_e$, and $\beta_e$ are, respectively, the electron
mass, energy, and velocity. 
The beautiful new experiments using traps to also measure with
high precision the recoil velocity of the daughter nucleus have observed
$b = -0.0027 \pm 0.0029$~\cite{ref:towner}, 
$a = 0.9981 {+0.0044 \atop -0.0048}$~\cite{Gorelov:2004hv}. 
The deviation of $a$ from the $V-A$ prediction $a=1$ 
leads to the (so far relatively weak) bound on the scalar mass
\beq
{M_S \over M_W} \sim (0.004)^{-1/4} \sim 4.
\eeq
A tighter bound of $(M_S/M_W) >6-7$ is expected from improved
measurements of $a$. Once the lower part of the beta spectrum is more
precisely measured, the overall normalization of the rate will yield a
more sensitive bound of $M_S > 15 M_W$ by utilizing interference of the
$S$ and $V-A$ amplitudes~\cite{ref:ashery}.

In passing, we note that standard
beta decay experiments such as KATERIN~\cite{ref:katrin},
which will measure the electron-neutrino mass (or rather $ m_{\nu_1}$)
down to 0.4~eV, will have very high statistics of $\sim 10 ^{11}$
events. Still, beta spectra with or without recoiling atoms are
also affected by radiative and hadronic effects, and precise
calculations of the latter will be required if the experimental
precision is to yield strong limits on non-standard couplings.

A scalar $\bar ud$ weak current
contributes to $G$-parity-violating second-class-current transitions,
such as \channel, provided that it couples to
$\tau^-$ and $\nu_\tau$.  As discussed above, the present experimental
upper bound on the branching ratio for this mode is an order of
magnitude greater than the estimated Standard-Model contribution $\sim
10^{-5}$, which is at the level that may be detected by the \babar\
and Belle experiments.
Since interference between a non-standard contribution and the small
$V-A$ amplitude will not contribute much, a limit of
the branching fraction at the level of $3\times 10^{-5}$ would imply
\beq
{M_S \over M_W} > (3\times 10^{-5})^{-1/4} \sim 12, 
\eeq
comparable to the expected future bounds from beta decay
experiments.
  
Unlike the universal gauged weak interactions, the scalar couplings
could discriminate between different lepton generations. Thus, the $S$
particle could be "first-generation oriented", coupling to the $u$ and
$d$ quarks and the $e$ and $\nu_e$ leptons but not to $\tau$ or
$\nu_\tau$. In such a case, it will affect the beta decays but not the
$\tau^- \to \pi^-\eta \nu_\tau$ decays. 
Conversely, $S$ particles may couple more strongly to the
third-generation $\tau\nu_\tau$ vertex than to $e\nu_e$.
%
%
Thus, {\it a-priori}, the limit from nuclear beta decays and the one
from the \channel\ decay are complementary and, furthermore, observation
of $S$-coupling effects in one mode and not the other would indicate
non-universality.

On the particle theory side, many lines of argument~\cite{Agashe:2004cp}
suggest that new physics, particularly novel weak couplings different from
standard $V-A$, will most strongly manifest in higher
generations. This would enhance $S$
effects in the $\tau$ decays relative to the first-generation 
beta decays.
More generally, $M_S$ is unlikely to
be much smaller than $M_P$, for which the very strict
bound above applies, unless $M_S$ is protected by $SU(2)_L$, namely,
$S$ couples to the $Z^0$. In that case, the $S^+$, $S^-$, and $S^0$ 
form an $SU(2)_L$ triplet, helping produce $S$ particles at the LHC via an
intermediate $Z^0$ or $W^\pm$. Otherwise, production of $S^+ S^-$ pairs
is smaller by $(\alpha_{EM}/{\alpha_{Weak}})^2 \sim 10^{-2}$.
In general, if we have left-right symmetry at relatively low
scales~\cite{Chang:1984uy} the stringent limits on $M_P$ push $M_S$ to
very high values.

\section{Conclusions}
\label{sec:VI}

We have considered the $S$- and $P$-wave contributions to the
branching fraction of the decay \channel. We find the $P$-wave
contribution, which is more robustly calculated, to be $3.6 \times
10^{-6}$, and the $S$-wave one to be around $1 \times 10^{-5}$, both in
agreement with previous calculations.  Given the capability of
experiments at the $B$~factories to measure or set a limit on the
branching fraction $\BR(\channel)$ at the $10^{-5}$ level, it is interesting to
note the implications of the possible experimental results:
\begin{itemize}
\item A "minimal" result of $\BR \sim (0.2-0.4)\times 10^{-5}$ with
the $\pi^- \eta$ invariant mass around the $\rho^-$ peak, which may be
hard to extract experimentally, involves no new surprises.
\item A larger value of $\BR$, in the range $(1-1.5)\times 10^{-5}$,
consistent with the chiral perturbation theory calculations and with our
quark-model estimate, would strongly suggest that $a_0^-(980)$
dominates the S-wave part of the decay. In this case, a narrow invariant-mass
peak around $980$~MeV should be seen.  This would strongly
suggest that the $a_0^- (980)$ is a $\bar u d$ scalar meson after all.
\item A somewhat larger value, $\BR > (2-3) \times 10^{-5}$ with
scalar-meson dominance, may indicate novel scalar components in the
weak interactions.
\end{itemize}

\begin{acknowledgments}
The authors thank Richard Kass for discussions that motivated
the present work, and Eric Braaten and Daniel Ashery for 
useful discussions.

\end{acknowledgments}

\end{document}